%% file: writeup.tex
\documentclass[fleqn,twoside]{article}
\usepackage{espcrc2,epsf}

\input{epsf2}

\usepackage{graphicx}
\usepackage[figuresright]{rotating}

\newcommand{\AmS}{{\protect\the\textfont2
  A\kern-.1667em\lower.5ex\hbox{M}\kern-.125emS}}

\newcommand{\mps}{m_{PS}}
\newcommand{\be}{\begin{equation}}
\newcommand{\ee}{\end{equation}}
\title{\vspace{-2.46em}
\parbox{\hsize}{\hbox to \hsize
{\hss \normalsize HU-EP-03/50}}
\parbox{\hsize}{\vspace{-0.7em}
\hbox to \hsize
{\hss \normalsize DESY 03-143}}
\parbox{\hsize}{\vspace{-1.2em}\hbox to \hsize
{\hss \normalsize LU-ITP 2003/016}}
\parbox{130mm}{\vspace{-7.8em}\hbox to \hsize
{\hss \normalsize LTH 590}}
\parbox{130mm}{\vspace{-8.2em}\hbox to \hsize
{\hss \normalsize Edinburgh 2003/17}}\\
\vspace{-3.3em}
Chiral Perturbation Theory and Finite Size Effects on the Nucleon Mass in 
unquenched QCD
\thanks{presented by A. Ali Khan}}

\author{A. Ali Khan\address[HU]{Institut f\"ur Physik, Humboldt-Universit\"at 
zu Berlin,  12489 Berlin, Germany}, 
T.~Bakeyev\address[DU]{Joint Institute for Nuclear Research, 141980 Dubna,
Russia},
M. G\"ockeler\address{Institut f\"ur Theoretische Physik, Universit\"at 
Leipzig, 04109 Leipzig, Germany}$^,$\address[RU]{Institut f\"ur Theoretische 
Physik, Universit\"at Regensburg, 93040 Regensburg, Germany}, 
T.R. Hemmert\address[TU]{Physik-Department, Theoretische Physik T39, TU M\"unchen,
85747 Garching, Germany},
R. Horsley\address[EU]{School of Physics, The University of 
Edinburgh, Edinburgh EH9 3JZ, UK}, 
A.C. Irving\address[LI]{Theoretical Physics Division, Department of 
Mathematical Sciences, University of Liverpool,  \\
Liverpool L69 3BX, UK},
D. Pleiter\address[NIC]{John von Neumann-Institut f\"ur Computing NIC, 
15738 Zeuthen, Germany}, 
P.E.L. Rakow\addressmark[LI], 
G. Schierholz\addressmark[NIC]$^,$\address[DESY]{Deutsches 
Elektronen-Synchrotron  DESY, 22603 Hamburg, Germany}, 
and H. St\"uben\address{Konrad-Zuse-Zentrum f\"ur Informationstechnik
Berlin, 14195 Berlin, Germany}
(QCDSF and UKQCD Collaborations)
}       
\begin{document}

\begin{abstract}
We calculate finite size effects on nucleon masses in chiral perturbation theory. We confront the
theoretical predictions with $N_f=2$ lattice results and discuss chiral extrapolation formulae.
\end{abstract}

\maketitle

\section{INTRODUCTION}
Finite size effects, in particular on dynamical configurations, can be a serious impediment
to precision lattice calculations of hadron masses and matrix elements. It is of interest to 
find a theoretical description of finite size effects, to be able to determine the
finite volume error at each lattice size and, possibly, to find an extrapolation formula to 
infinite volume. 

On the lattice sizes used in production runs, the nucleons and pions are the
relevant degrees of freedom  for understanding the finite size effects on the nucleon mass. 
To calculate them, we use  two-flavor relativistic baryon chiral perturbation theory ($\chi PT$) at $O(q^3)$ in the chiral counting (one-loop order).
\section{LATTICE PARAMETERS}
\begin{figure}[t]
\epsfysize=6.7cm \epsfbox{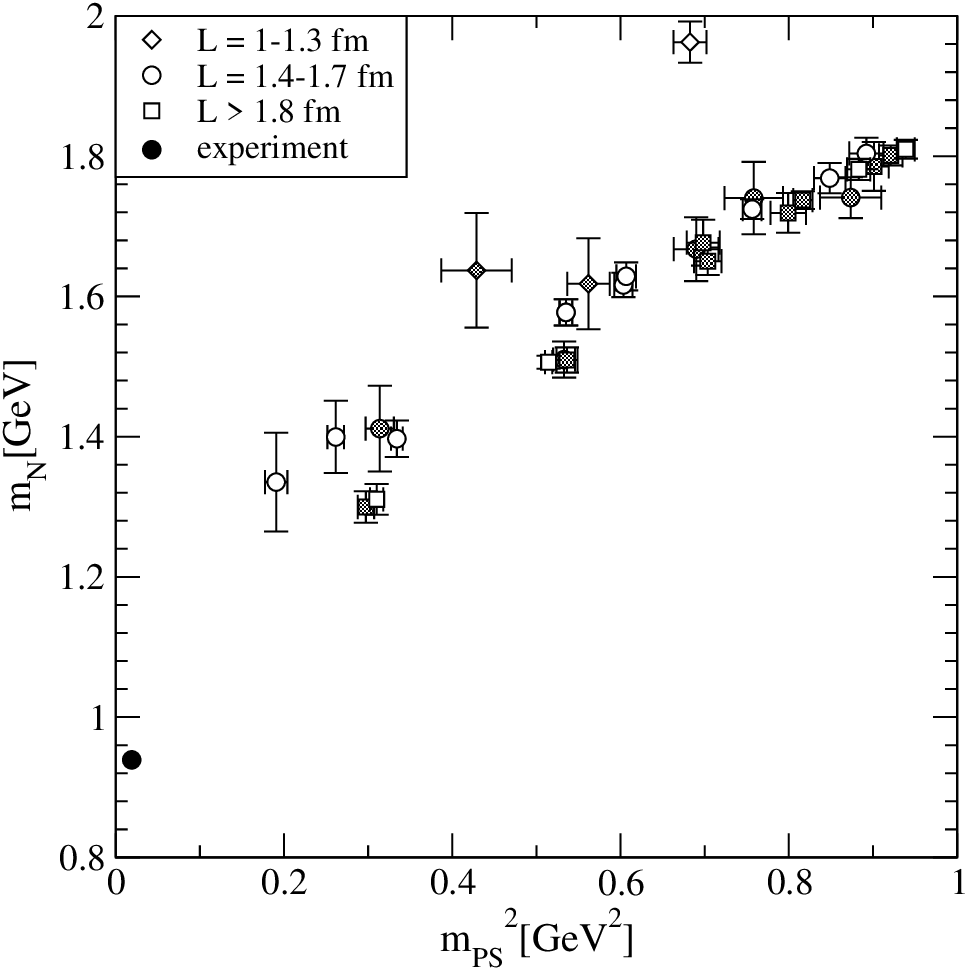}
\vspace{-20pt}
\caption{Nucleon masses from UKQCDSF (white symbols) and JLQCD (gray symbols).
}
\label{fig:oaimproved}
\vspace{-20pt}
\end{figure}

We base our finite volume study on QCDSF and UKQCD configurations, 
using a plaquette gauge action with two flavors of non-perturbatively $O(a)$-improved Wilson fermions ($a$ denotes the lattice spacing).
The pion masses are in the interval $0.4-1$ GeV. 
Valence and sea quark masses are equal. Lattices volumes are $1-2.2$ fm.
The scale is set with $r_0$, using the value at the physical point
$r_0 = 0.5$ fm $\simeq 1/(395 \mathrm{MeV}$). 
We compare our results to JLQCD  data with
the same lattice actions and range of simulation 
parameters on varying lattice sizes~\cite{JLQCD02}. 
In Fig.~\ref{fig:oaimproved} we plot nucleon masses of both groups.
It is found that the masses increase when the lattice size is decreased.
The data from lattices of an extent $\geq 1.8$ fm lie approximately on the
same curve. In principle there are two major sources of systematic error: finite
volume and discretization effects. Among the data points, the lattice spacing  
varies between $0.4$ and $0.6 \;\mathrm{GeV}^{-1}$. The $a \rightarrow 0$ limit 
was not performed, and we have to ascertain that there are no sizable 
discretization errors in the results. 
For example, between two data points on approximately the same volume and at the same 
pion mass, but with $a$ varying by $\sim 30\%$, the nucleon mass is found to remain 
unchanged within the statistical errors. 
We also compare our large lattice results ($L \geq 1.8$ fm) 
with two CP-PACS data sets~\cite{cppacs01} using 
renormalization group improved gauge fields and tree-level tadpole-improved clover quarks
 at $a^{-1}\approx 1.5$ and 2 GeV respectively. Their box sizes are $\geq 2.5$ fm.
A compilation of the large $L$ data is given in Fig.~\ref{fig:chPT}. We find a good scaling and 
conclude that in this data $O(a)$ uncertainties  are small.
%
%
\section{CHIRAL EXTRAPOLATION}

The one-loop contribution is generated by the $O(q^1)$ Lagrangian ${\cal L}_N^{(1)}$:
\be
{\cal L}^{(1)}_N =  \bar{\Psi}\left(i\gamma_\mu D^\mu - m_0\right)\Psi + 
\frac{1}{2} g_A\bar{\Psi} \gamma_\mu \gamma_5 u^\mu \Psi ,
\ee
with $D_\mu = \partial_\mu + \frac{1}{2}[u^\dagger,\partial_\mu u]$, 
$u_\mu = iu^\dagger \partial_\mu U u^\dagger$ and $u^2 = U$.

\begin{figure}[tb]
\vspace{-2pt}
\epsfysize=7.0cm \epsfbox{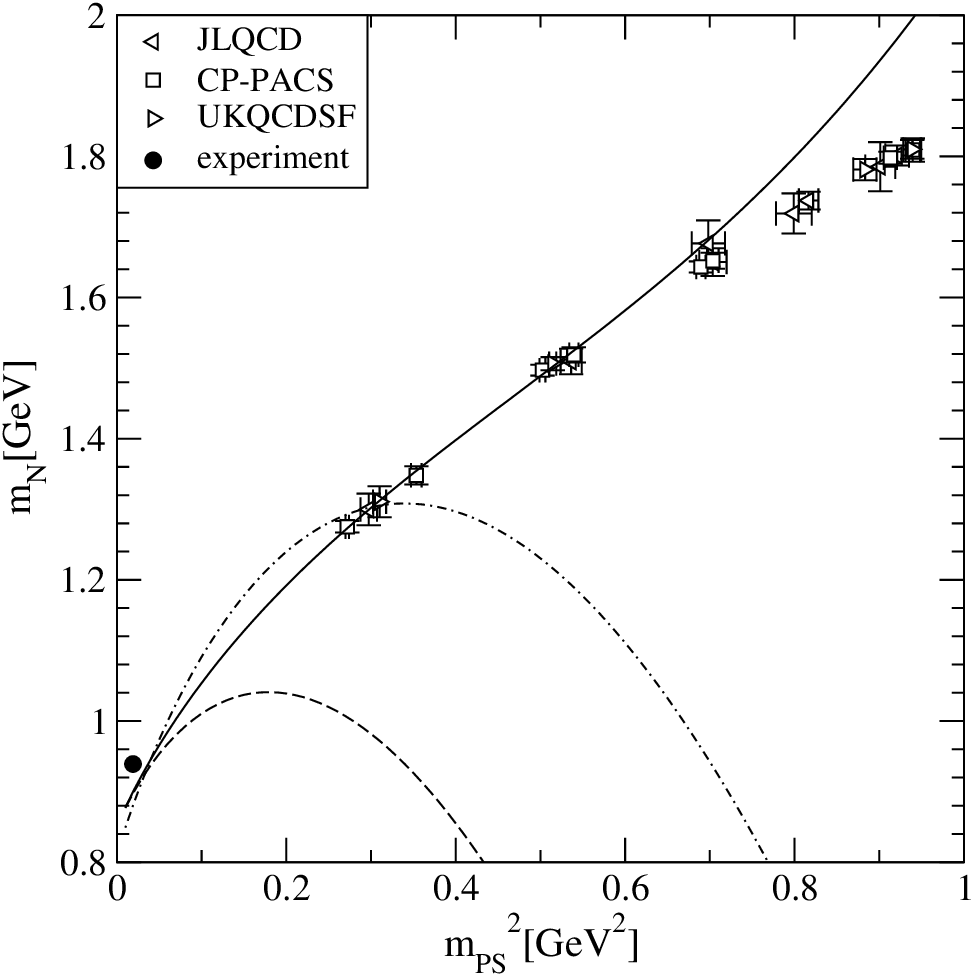}
\vspace{-20pt}
\caption{Nucleon mass compared to $\chi PT$. The solid curve denotes
the fit with relativistic $\chi PT$, and the dashed curve the non-relativistic limit using
the same values of $m_0$ and $c_1$. 
The dot-dashed curve shows a non-relativistic
result with estimated parameters, $m_0 = 0.81$ GeV and $c_1 = -1.1$ GeV$^{-1}$.
}
\vspace{-15pt}
\label{fig:chPT}
\end{figure}

We use the infrared regularization 
 scheme which is discussed in detail in~\cite{becher99}.
To compute the renormalized nucleon mass $m_N$, we add at tree-level the ${\cal L}^{(2)}_N$ term 
$-4c_1\mps^2$ and an additional term of the form $e_1\mps^4$, which is, 
strictly speaking, derived from  ${\cal L}^{(4)}_N$.
The renormalization procedure is detailed in~\cite{proc_future}. 
In this calculation we use $g_A =  1.2$~\cite{proc03}, and $F = 92.4$ MeV.
We determine the nucleon mass in the chiral limit, $m_0$,  the value of $c_1$ and the
renormalized $e_1$ ($e_1^r$) by a fit to six lattice data points at the smallest
masses, and find
$m_0 = 0.85(14)\mathrm{GeV}$, $c_1=-0.80(18)\mathrm{GeV}^{-1}$ and $e_1^r(1
\,\mathrm{GeV})=2.8(1.1) \mathrm{GeV}^{-3}$. For the definition of $e_1^r$
see \cite{proc_future}.
In Fig.~\ref{fig:chPT}, the result is compared with the lattice data on volumes $> 1.8$ fm.
The $\chi PT$ result shown here differs numerically slightly  from the one quoted 
in~\cite{proc_future} since they used the value $g_A=1.267$ and  a fit 
to a larger set of lattice points.
Expanding the $\chi PT$ result up to $O(\mps^3)$,
one obtains the non-relativistic (NR) approximation. The NR theory is valid only
in the limit $\mps \ll m_N$. This is reflected in the breakdown of the curves already at small 
pion masses, which is also shown in Fig.~\ref{fig:chPT}. A method to push the validity of the NR 
approximation  to higher momentum scales within a cutoff scheme is described in~\cite{bernard03}.
\section{FINITE SIZE EFFECTS}
We calculate the finite size effect from the one-loop $O(q^3)$ contribution 
to the self energy. Putting the external nucleon line on-shell,
 this is given by~\cite{becher99}
\begin{eqnarray}
 \Sigma(q\!\!\slash=m_0) & = & 
-i\frac{3g_A^2m_0\mps^2}{2F^2}\int_0^\infty dx \nonumber \\
 & & \!\!\!\!\!\!\!\!\!\!\!\!\!\!\!\!\!\!\!\!\!\!\!\!
\!\!\!\!\!\!\!\!\!\!\!\!\!\!\!\!\int\frac{d^4p}{(2\pi)^4}
\left[p^2-m_0^2x^2 - \mps^2(1-x)+i\epsilon\right]^{-2}  
\end{eqnarray}
in Minkowski space. We define
\begin{eqnarray} 
 \delta  &=& \frac{1}{m_0}
\left(\Sigma(q\!\!\slash=m_0,L) - \Sigma(q\!\!\slash=m_0,\infty)\right)
\end{eqnarray}
where the temporal extent of the lattice is assumed to be infinite.
Using~\cite{hasenfratz90}, one can express $\delta$ as an integral over 
Bessel functions:
\begin{eqnarray}
\delta &=& \frac{3g_A^2\mps^2}{16\pi^2F^2}
  \int_0^\infty dx \times  \nonumber \\ & \times &  \sum_{\vec{n}\neq 0} 
 K_{0}\left(L|\vec{n}|\sqrt{m_0^2x^2+\mps^2(1-x)}\right).
\end{eqnarray}
To calculate the nucleon mass in a finite box,
\be
 m_N(L) =  (1+\delta)m_N(\infty), \label{eq:delta}
\ee
we first extrapolate to $m_N(\infty)$ by using the lattice result on the largest 
available lattice as input to~Eq.(\ref{eq:delta}). We are then able to calculate $m_N(L)$ at 
finite lattice extent.
\begin{figure}[thb]
\epsfysize=6.7cm \epsfbox{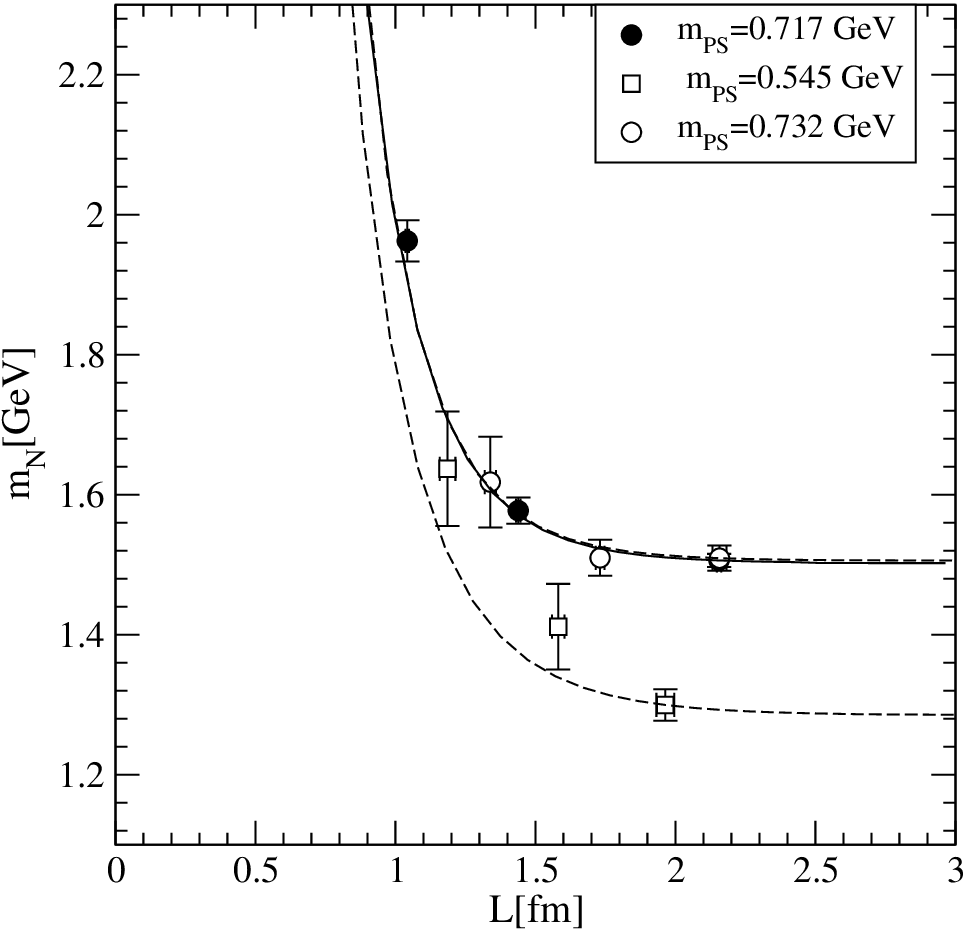}
\vspace{-20pt}
\caption{Comparison of the finite size dependence of lattice nucleon mass
data sets at a fixed $(\beta,\kappa)$ value and $\chi PT$. The data sets 
are labeled by the pion mass on the largest volume.}
\label{fig:chPTfinitevol}
\vspace{-15pt}
\end{figure}
Lattice results for the nucleon mass at fixed values of $\beta$ and
$\kappa$, but different lattice sizes are compared with chiral
perturbation theory in Fig.~\ref{fig:chPTfinitevol}. It is found that the finite size effect
can be well described by relativistic $\chi PT$.
In contrast, in NR $\chi PT$ at $O(q^3)$
we obtain  only roughly $\sim 30\%$ of the finite volume 
effect in the lattice data~\cite{alikhan02}. At pion masses $\geq 500$ MeV, large
loop momenta give a substantial contribution also to the finite size effect. 
It is of
interest to study whether this dependence is reduced in the relativistic formalism.

\noindent {\bf Acknowledgements} \\
This work is supported in part by the Deutsche Forschungsgemeinschaft. 
The simulations were done on the APEmille at NIC (Zeuthen),
  the Hitachi SR8000 at LRZ (Munich) and the Cray T3E at EPCC (Edinburgh)
  and NIC (J\"{u}lich). We thank all institutions for their support.

\end{document}

%% file: epsf2.tex
\ifx\epsfannounce\undefined \def\epsfannounce{\immediate\write16}\fi
 \epsfannounce{This is `epsf.tex' v2.7k <10 July 1997>}%
\newread\epsffilein    
\newif\ifepsfatend     
\newif\ifepsfbbfound   
\newif\ifepsfdraft     
\newif\ifepsffileok    
\newif\ifepsfframe     
\newif\ifepsfshow      
\epsfshowtrue          
\newif\ifepsfshowfilename 
\newif\ifepsfverbose   
\newdimen\epsfframemargin 
\newdimen\epsfframethickness 
\newdimen\epsfrsize    
\newdimen\epsftmp      
\newdimen\epsftsize    
\newdimen\epsfxsize    
\newdimen\epsfysize    
\newdimen\pspoints     
\def\epsfbox#1{\global\def\epsfllx{72}\global\def\epsflly{72}%
   \global\def\epsfurx{540}\global\def\epsfury{720}%
   \def\lbracket{[}\def\testit{#1}\ifx\testit\lbracket
   \let\next=\epsfgetlitbb\else\let\next=\epsfnormal\fi\next{#1}}%
\def\epsfgetlitbb#1#2 #3 #4 #5]#6{%
   \epsfgrab #2 #3 #4 #5 .\\%
   \epsfsetsize
   \epsfstatus{#6}%
   \epsfsetgraph{#6}%
}%
\def\epsfnormal#1{%
    \epsfgetbb{#1}%
    \epsfsetgraph{#1}%
}%
\newhelp\epsfnoopenhelp{The PostScript image file must be findable by
TeX, i.e., somewhere in the TEXINPUTS (or equivalent) path.}%
\def\epsfgetbb#1{%
    \openin\epsffilein=#1
    \ifeof\epsffilein
        \errhelp = \epsfnoopenhelp
        \errmessage{Could not open file #1, ignoring it}%
    \else                       
        {
            \chardef\other=12
            \def\do##1{\catcode`##1=\other}%
            \dospecials
            \catcode`\ =10
            \epsffileoktrue         
            \epsfatendfalse     
            \loop               
                \read\epsffilein to \epsffileline
                \ifeof\epsffilein 
                \epsffileokfalse 
            \else                
                \expandafter\epsfaux\epsffileline:. \\%
            \fi
            \ifepsffileok
            \repeat
            \ifepsfbbfound
            \else
                \ifepsfverbose
                    \immediate\write16{No BoundingBox comment found in %
                                    file #1; using defaults}%
                \fi
            \fi
        }
        \closein\epsffilein
    \fi                         
    \epsfsetsize                
    \epsfstatus{#1}%
}%
\def\epsfclipoff{\def\epsfclipstring{\ifepsfdraft\space clip\fi}}%
\epsfclipoff 
\def\epsfspecial#1{%
     \epsftmp=10\epsfxsize
     \divide\epsftmp\pspoints
     \ifnum\epsfrsize=0\relax
       \includegraphics{\ifepsfdraft}%
     \else
       \epsfrsize=10\epsfysize
       \divide\epsfrsize\pspoints
       \includegraphics{\ifepsfdraft}%
     \fi
}%
\def\epsfframe#1%
{%
  \leavevmode                   
  \setbox0 = \hbox{#1}%
  \dimen0 = \wd0                                
  \advance \dimen0 by 2\epsfframemargin         
  \advance \dimen0 by 2\epsfframethickness      
  \vbox
  {%
    \hrule height \epsfframethickness depth 0pt
    \hbox to \dimen0
    {%
      \hss
      \vrule width \epsfframethickness
      \kern \epsfframemargin
      \vbox {\kern \epsfframemargin \box0 \kern \epsfframemargin }%
      \kern \epsfframemargin
      \vrule width \epsfframethickness
      \hss
    }
    \hrule height 0pt depth \epsfframethickness
  }
}%
\def\epsfsetgraph#1%
{%
   \leavevmode
   \hbox{
     \ifepsfframe\expandafter\epsfframe\fi
     {\vbox to\epsfysize
     {%
        \ifepsfshow
            \vfil
            \hbox to \epsfxsize{\epsfspecial{#1}\hfil}%
        \else
            \vfil
            \hbox to\epsfxsize{%
               \hss
               \ifepsfshowfilename
               {%
                  \epsfframemargin=3pt 
                  \epsfframe{{\tt #1}}%
               }%
               \fi
               \hss
            }%
            \vfil
        \fi
     }%
   }}%
   \global\epsfxsize=0pt
   \global\epsfysize=0pt
}%
\def\epsfsetsize
{%
   \epsfrsize=\epsfury\pspoints
   \advance\epsfrsize by-\epsflly\pspoints
   \epsftsize=\epsfurx\pspoints
   \advance\epsftsize by-\epsfllx\pspoints
   \epsfxsize=\epsfsize{\epsftsize}{\epsfrsize}%
   \ifnum \epsfxsize=0
      \ifnum \epsfysize=0
        \epsfxsize=\epsftsize
        \epsfysize=\epsfrsize
        \epsfrsize=0pt
      \else
        \epsftmp=\epsftsize \divide\epsftmp\epsfrsize
        \epsfxsize=\epsfysize \multiply\epsfxsize\epsftmp
        \multiply\epsftmp\epsfrsize \advance\epsftsize-\epsftmp
        \epsftmp=\epsfysize
        \loop \advance\epsftsize\epsftsize \divide\epsftmp 2
        \ifnum \epsftmp>0
           \ifnum \epsftsize<\epsfrsize
           \else
              \advance\epsftsize-\epsfrsize \advance\epsfxsize\epsftmp
           \fi
        \repeat
        \epsfrsize=0pt
      \fi
   \else
     \ifnum \epsfysize=0
       \epsftmp=\epsfrsize \divide\epsftmp\epsftsize
       \epsfysize=\epsfxsize \multiply\epsfysize\epsftmp
       \multiply\epsftmp\epsftsize \advance\epsfrsize-\epsftmp
       \epsftmp=\epsfxsize
       \loop \advance\epsfrsize\epsfrsize \divide\epsftmp 2
       \ifnum \epsftmp>0
          \ifnum \epsfrsize<\epsftsize
          \else
             \advance\epsfrsize-\epsftsize \advance\epsfysize\epsftmp
          \fi
       \repeat
       \epsfrsize=0pt
     \else
       \epsfrsize=\epsfysize
     \fi
   \fi
}%
\def\epsfstatus#1{
   \ifepsfverbose
     \immediate\write16{#1: BoundingBox:
                  llx = \epsfllx\space lly = \epsflly\space
                  urx = \epsfurx\space ury = \epsfury\space}%
     \immediate\write16{#1: scaled width = \the\epsfxsize\space
                  scaled height = \the\epsfysize}%
   \fi
}%
{\catcode`\%=12 \global\let\epsfpercent=
\global\def\epsfatend{(atend)}%
\long\def\epsfaux#1#2:#3\\%
{%
   \def\testit{#2}
   \ifx#1\epsfpercent           
       \ifx\testit\epsfbblit    
            \epsfgrab #3 . . . \\%
            \ifx\epsfllx\epsfatend 
                \global\epsfatendtrue
            \else               
                \ifepsfatend    
                \else           
                    \epsffileokfalse
                \fi
                \global\epsfbbfoundtrue
            \fi
       \fi
   \fi
}%
\def\epsfempty{}%
\def\epsfgrab #1 #2 #3 #4 #5\\{%
   \global\def\epsfllx{#1}\ifx\epsfllx\epsfempty
      \epsfgrab #2 #3 #4 #5 .\\\else
   \global\def\epsflly{#2}%
   \global\def\epsfurx{#3}\global\def\epsfury{#4}\fi
}%
\def\epsfsize#1#2{\epsfxsize}%